\documentclass[fleqn,usenatbib]{mnras}
\usepackage[T1]{fontenc}
\usepackage{ae,aecompl}
\usepackage{latexsym,amssymb}
\usepackage{graphicx}
\usepackage{ulem}

\newcommand{\msun}{\mbox{$M_\odot$}}

\title[Disk formation and jet inclination effects in Common Envelopes]{Disk formation and jet inclination effects in Common Envelope}

\author[L\'opez-C\'amara et al.]{Diego L\'opez-C\'amara$^{1}$\thanks{E-mail: diego@astro.unam.mx}
, Enrique Moreno M\'endez$^{2}$
, Fabio De Colle$^{3}$
\\
$^{1}$CONACyT - Instituto de Astronom\'ia, Universidad Nacional Aut{\'o}noma de M{\'e}xico, A. P. 70-264 04510 CDMX, Mexico\\
$^{2}$Facultad de Ciencias, Universidad Nacional Aut{\'o}noma de M{\'e}xico, A. P. 70-543 04510 D. F. Mexico\\
$^{3}$Instituto de Ciencias Nucleares, Universidad Nacional Aut\'onoma de M\'exico, A. P. 70-543 04510 D. F. Mexico}
\date{Accepted XXX. Received YYY; in original form ZZZ}

\pubyear{2020}

\begin{document}
\label{firstpage}
\pagerange{\pageref{firstpage}--\pageref{lastpage}}
\maketitle

\begin{abstract}
The evolution and physics of the common envelope (CE) phase are still not well understood. Jets launched from a compact object during this stage may define the evolutionary outcome of the binary system. We focus on the case in which jets are launched from a neutron star (NS) engulfed in the outer layers of a red giant (RG). We run a set of three-dimensional hydrodynamical simulations of jets with different luminosities and inclinations. The luminosity of the jet is self-regulated by the mass accretion rate and an efficiency $\eta$. Depending on the value of $\eta$ the jet can break out of the BHL bulge (``successful jet") and aligns against the incoming wind, in turn, it will realign in favour of the direction of the wind. The jet varies in size and orientation and may present quiescent and active epochs. The inclination of the jet and the Coriolis and centrifugal forces, only slightly affect the global evolution. As the accretion is hypercritical, and the specific angular momentum is above the critical value for the formation of a disk, we infer the formation of a disk and launching of jets. The disks' mass and size would be $\sim$10$^{-2}$~M$_\odot$ and $\gtrsim 10^{10}$ cm, and it may have rings with different rotation directions. In order to have a successful jet from a white dwarf, the ejection process needs to be very efficient ($\eta\sim$0.5). For main sequence stars, there is not enough energy reservoir to launch a successful jet.
\end{abstract}

\begin{keywords}
Binaries: general  --
Binaries (including multiple): close
Accretion, accretion disks --
Stars: jets 
Methods: numerical -- 
Hydrodynamics --
\end{keywords}

\section{Introduction}
\label{sec:int}
The common envelope (CE) phase is an important stage of binary systems. Even though it is short in span ($t\lesssim 10^2-10^4$~yr; see, e.g., \citealt{meyer79, igoshev19}), it is a key part of the evolutionary channels which lead to transitory and highly energetic astrophysical phenomena (THEAP), for example, some supernovae (SNe) \citep[][]{chevalier12, py14}, long gamma-ray bursts (GRBs) \citep[][]{Brown2007,rr09}, short GRBs \citep[][]{vg20} and gravitational waves \citep[][]{abbott16}.

During a CE phase one of the stars in a binary engulfs its companion \citep{paczynski76}. Due to mass transfer and orbital decay (because of energy and angular momentum loss) the orbital separation decreases up to a point in which a THEAP may be produced (we refer the reader to \citealt{ivanova13} for a review of the CE phase). In this paper, we focus on the case in which the star is a massive red giant (RG) and the engulfed member is a neutron star (NS). As \citet{py14} point out, this may be the result of the evolution of two massive ZAMS stars (10-20 M$_\odot$ and 10-16 M$_\odot$) in a binary system in which the most massive of the two becomes a RG and transfers several solar masses to the companion through Roche Lobe overflow \citep[RLOF,][]{jones20}, or through wind-RLOF \citep{mp07, elmellah19} and explodes as a SN leaving behind a NS. A key point in the evolution of this particular system is that the SN does not break the binary system. Later, the companion becomes a RG, expands beyond its Roche Lobe (RL), and engulfs the NS in a CE. 

The CE has been followed by many analytical \citep{rasio91, iben93, han94, terman94, regos95, armitage00, sandquist00, papish15, macleod17b}, as well as numerical studies

\newpage
\noindent \citep{taam78, meyer79, bodenheimer84, livio88, rasio96, sandquist98, demarco03, lombardi06, ricker08, taam10, demarco11, passy11, passy12, ricker12, nandez14, macleod15a, macleod15b, ivanova16, kuruwita16, nandez16, ohlmann16a, ohlmann16b, pejcha16, staff16, hillel17, iaconi17, macleod17a, murguia17, ohlmann17,chamandy18, grichener18, chamandy19a, chamandy19b, fragos19, grichener19, iaconi19a, iaconi19b, prust19, reichardt19a, reichardt19b, glanz20, prust20}, nevertheless, many open questions remain unanswered. The dominant processes in the evolution and termination of the CE, for example, are not yet fully understood, and many mechanisms have been proposed (e.g., the $\alpha_{CE}-\lambda$ formalism, \citealt{heuvel76}; gamma formalism, \citealt{nelemans00}; recombination energy, \citealt{nandez15}; binding energy, \citealt{iaconi18}; tidal forces, \citealt{iben93}; internal energy, \citealt{han94}; magnetic fields, \citealt{regos95}; accretion, \citealt{voss03, soker04}; nuclear  energy, \citealt{ivanova03}; dust-driven winds \citealt{glanz18}; and enthalpy, \citealt{ivanova11}, amongst others), none of which have been able to fully describe the evolution and termination of the CE phase \citep[see for example][]{soker13a, iaconi18}. The launching of a jet from a compact object (CO) within the CE is also a possible alternative mechanism in the evolution and termination of this phase \citep{armitage00, soker04, chevalier12, shiber16, soker18, gilkis19, schreier19, soker19, jones20}.

Numerical hydrodynamical (HD) studies of jets within the CE (or in the outskirts) have also been performed {\citep{soker13b, soker14, mmlcdc17, sabach17, shiber17, shiber18, lcdcmm19, shiber19}}. These show that the effects that jets produce in the CE may be substantial, and may be able to unbind three times as much envelope mass than when no jets are present {\citep{shiber19}}. We must note that the interaction between the expanding cocoon (formed as a result of the interaction between the jet and the environment) and the CE material would reduce the accretion rate onto the CO. This is termed ``negative jet feedback" \citep[NJF,][]{soker16, soker18}, and has been confirmed by numerical simulations \citep{mmlcdc17, lcdcmm19}. 

None of the mentioned HD jet studies have focused on whether the accreted mass can form a disk. In order to launch a jet within the CE phase an accretion disk around the stellar companion is necessary. When the companion is a CO, it is likely that the angular momentum of the RL-transferred material will be enough for a disk to form around it. This process, nevertheless, is still not fully understood. Some authors claim that no accretion disk would form due to the low mass accretion rates obtained \citep{macleod15a, macleod15b}, while others find that the mass accretion rate is large enough to possibly produce an accretion disk \citep{shiber16, chamandy18, grichener18}.

Phases which involve catastrophic mass-loss events (e.g. core collapse SNe, GRBs) may misalign the spins of the binary system members through SNe kicks \citep{spruit98, wongwa13, enrique16}. An alternative route is to consider that the binary system formed in dense clusters, where stellar-companion captures are more likely. In such cases, the orientation of the angular momentum of the COs and the orbit may be misaligned \citep{banerjee18}. Interactions in triple stellar systems \citep{schreier19}, as well as the Lense-Thiring and the Bardeen-Peterson effects \citep[][respectively]{lt18, bp75}, may also produce misaligned jets. Thus, if one of the member of the binary system were to launch jets within the CE, they may do so with an inclination angle (with respect to the orbital plane).

In this study we follow the interaction between the material that is accreted by the NS, the incoming wind material (due to the orbital motion of the NS within the CE and that we are in the comoving frame), and the jet which is launched from the NS. We run a set of three-dimensional (3D) HD simulations of jets launched with different luminosities and inclination angles. The luminosity is self-regulated according to the mass accretion rate and an efficiency $\eta$. We focus on the mass accretion rate and specific angular momentum in order to analyse the possible formation of an accretion disk. Coriolis and centrifugal forces are included, and we follow the models up to approximately a week.

The paper is organised as follows. In Section~\ref{sec:setup} we describe the numerical setup, and the characteristics of each model. In Section~\ref{sec:global_mdot}, we discuss the global evolution, and the mass accretion rate of the vertical jet models. In Section~\ref{sec:incl_disk}, we analyse the effects of inclined jets and the possible formation of a disk. In Section~\ref{sec:con}, we present our conclusions.

\section{Setup and models}
\label{sec:setup}
We follow the evolution of a self-regulated jet launched from a NS, moving within the envelope of a massive RG in the CE phase using 3D-HD numerical simulations using the code {\it Mezcal} \citep{decolle12a}. 

The initial setup consists of the stellar envelope of a $20$-M$_\odot$ RG with a mass profile given by \citet{papish15} and a $1.4$-M$_\odot$ NS orbiting around the centre of mass of the NS-RG system. Unless stated, the setup is akin to that from \citet{lcdcmm19}. We assume a system where no tidal synchronization has occurred, thus the NS moves with Keplerian velocity ($v_{\rm{k}}$) with respect to the CE. The numerical domain covers approximately $\sim$10\% of the CE of the RG (R$_{\rm{RG}}=3.7\times10^{13}$~cm) and the orbital separation between the NS and the core of the RG is $a=1.1\times 10^{13}$~cm. We model only the top half of the system. We set the reference system on the comoving frame of the NS, with which the orbital motion of the NS around the RG is reflected by a wind with velocity $v_{\rm w}=-v_{\rm k}$ which is incoming towards the NS. The gravitational effects of the NS were modelled by considering a $1.4$~M$_\odot$ point mass located at (0,0,$a$).

The jet is launched from a spherical inner boundary ($r_{\rm{in}}=10^{11}$~cm) with an inclination angle $\theta_i$ (measured from the polar axis). The luminosity of the jet is self-regulated by the mass accretion rate $\dot{M}_{\rm{r,in}}$ and an efficiency fraction $\eta$ between 0 and 1 (this is, $L_{\rm{j}} = \eta \dot{M}_{\rm{r,in}} {\rm{c}}^2$). The opening angle and velocity of the jet are set to 15$^\circ$ and $v_{\rm{j}}=c/3$. We keep track of the mass accretion rate and the specific angular momentum of the material that crosses $r_{\rm{in}}$ ($\dot{M}_{\rm{r, in}}$ and $\dot{J}_{\rm{r, in}}$, respectively). Centrifugal and Coriolis forces are included in the calculations. Since general relativistic effects, magnetic (B) field effects, and the self-gravity of the stellar material are negligible (compared to the gravitational, centrifugal and Coriolis forces, and to the ram pressure of the jet, they are not considered. The equatorial plane of the computational domain covers $x_{\rm{min,max}}=(-2\times10^{12}$, $2\times10^{12}$)~cm and $z_{\rm{min,max}}=(9\times10^{12}$, $13\times10^{12}$)~cm, while the polar axis covers $y_{\rm{min,max}}=(0$, $4\times10^{12}$)~cm (``small" domain). In some models, $x_{\rm{max}}$ was extended up to $6\times10^{12}$~cm (``big" domain). The equatorial plane has a reflective boundary, except for the jet launching region within $r_{\rm{in}}$. The wind is injected from the $zy$ boundary plane at $x=x_{\rm{max}}$. All other boundaries are set with outflow conditions. Six levels of refinement ranging from $7.8125 \times 10^{9}$~cm to $2.5 \times 10^{11}$~cm are used. The total integration time for all models is 5.51$\times 10^{5}$~s.

We follow a set of self-regulating jet models with different $\eta$ and $\theta_i$ values (and different domain sizes). We label the models according to whether the jet is launched parallel (``p") or inclined (``i") with respect to the polar axis; the value of $\eta$; and whether the domain size is ``small'' or ``big'' (``s" or ``b", respectively). If the jet is inclined, we also include the angle with which the jet is inclined (in the $XY$ plane). For instance, model i0.02b+45 refers to a jet inclined +45$^\circ$, with an efficiency $\eta =2\%$, and which is followed in the ``big'' domain. The label, inclination angle ($\theta_i$), the plane in which the jet is launched in, the efficiency $\eta$, and domain size of each of the models is shown in Table~\ref{table1}.

\begin{table}
\caption{Initial conditions of the numerical models}
\begin{center}
\begin{tabular}{ccccc}
  \hline
  Model &    $\theta_i$      & Plane  &   $\eta$ &  Domain   \\
             &   ($^\circ$)     &            &               &                 \\  
  \hline
  p0.00s  &  0   &     -     & 0.00 & small  \\
  p0.001s  &  0   &     -     & 0.001 & small  \\
  p0.01s  &  0   &     -     & 0.01 & small  \\ 
  p0.02s  &  0   &     -     & 0.02 & small  \\  
  p0.05s  &  0   &     -     & 0.05 & small  \\  
  p0.05b  &  0   &     -     & 0.05 & big     \\     
  i0.02b-45    & -45  & XY  & 0.02 & big  \\
  i0.02b+45   & +45  & XY  & 0.02 & big  \\      
  \hline
\end{tabular}
\end{center}
\label{table1}
\end{table}

\section{Global evolution and mass accretion rates}
\label{sec:global_mdot}
\subsection{Global evolution}
\label{sec:global}
In this section, we describe the outcome of the 3D-HD simulations. The presence of the NS in the envelope of the RG produces Bondi-Hoyle-Littleton accretion \citep[BHL,][]{bh44} very much alike that shown in \citet[][]{mmlcdc17, lcdcmm19}. Due to the BHL accretion, a bulge (termed BHL bulge) is formed, which completely engulfs the NS. Centrifugal and Coriolis forces twist the BHL bulge towards the stellar core.

\begin{figure*}
   \includegraphics[width=0.9\textwidth]{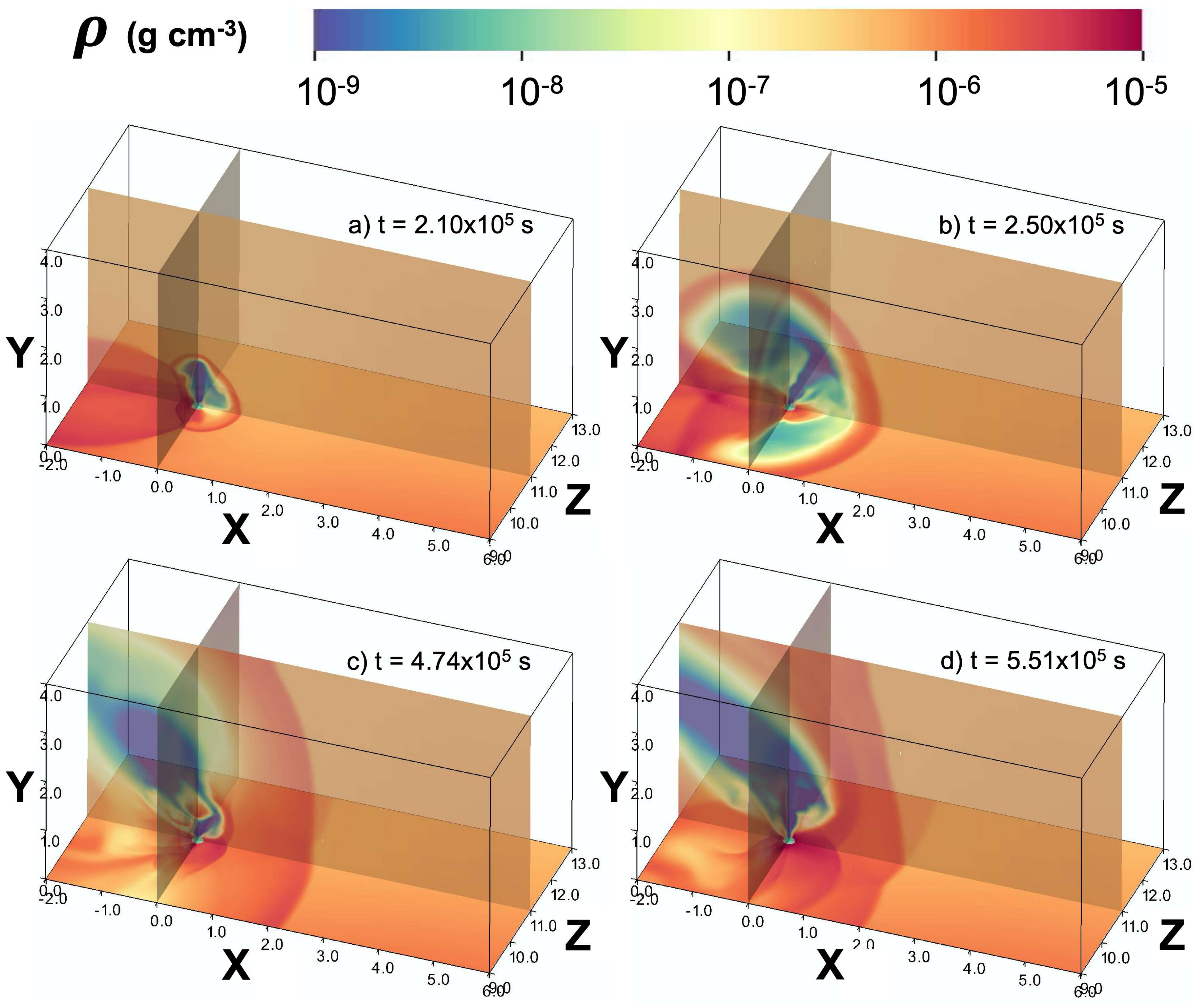}
   \caption{Density (in g~cm$^{-3}$) map slices ($XZ$, $XY$, and $YZ$) showing the global evolution of model p0.05b (at $2.10\times 10^{5}$~s, $2.50 \times 10^{5}$~s, $4.74 \times 10^{5}$~s, and $5.51 \times 10^{5}$~s, from top left to bottom right respectively). The axis are in units of $10^{12}$~cm.}
   \label{fig1}
\end{figure*}

Once the system achieves a quasi-steady mass accretion rate, the jet is launched from the spherical radius $r_{\rm{in}}$ (at t=2$\times$10$^{5}$~s). For our setup, we find that models with enough ram pressure ($\eta \gtrapprox 0.02$) are able to drill through the BHL bulge producing ``successful" jets, while in models with smaller ram pressure efficiency jets are choked. An interesting feature in low-$\eta$ models is that the jet may be able to break out of the bulge and then may be forever quenched{\footnote{The jet from model p0.01s, for example, breaks out of the bulge and 3.4$\times$10$^{4}$~s after its launch is chocked for the rest of the integration time.}}. The dynamical evolution of the successful jets can be described as follows (see Figure~\ref{fig1} showing density map slices for model p0.05b):

i) The self-regulated jet is able to drill through the BHL bulge and is subject to the NJF mechanism (which is consequence of the unsteady mass accretion rate onto the NS \citealt{soker16}). Due to the NJF, the jet power, size, and direction varies as a function of time. Panel a of Figure~\ref{fig1} shows how the jet with $\eta=0.05$ (model p0.05b) is able to break out of the bulge. The jet has a density of order $\sim10^{-9}$~g~cm$^{-3}$. During its propagation, the jet remains mostly vertically aligned, and reaches $y\sim$10$^{12}$~cm after $\sim$10$^{4}$~s. The environment and the jet material are shocked by the forward and reverse jet shocks, respectively. Hot material in the post-shock region expands laterally forming an extended cocoon. The cocoon for model p0.05b (with $\sim 10^{-8}-10^{-7}$~g~cm$^{-3}$), expands smoothly over the stellar envelope of the RG and reaches a radius of $\sim$10$^{12}$~cm.

ii) For the successful jet models, the jet is first pushed against the wind due to the pressure of the bulge, then the jet is pushed versus the incoming wind (i.e. in the laboratory frame the jet will align in favour with the direction of the orbital motion of the NS). As this happens, the jet continues presenting the variability in power and direction mentioned in i). Meanwhile, the cocoon expands smoothly over the domain. Panel b of Figure~\ref{fig1} shows how the jet of model p0.05b is pushed against the wind and how the cocoon has reached $y\sim3\times 10^{12}$~cm and $x\sim 1.5\times 10^{12}$~cm.

iii) At some moment, due to the winds ram pressure, the successful jets are realigned with the wind (against the direction of the orbital motion of the NS in the laboratory frame). This is due to the ram pressure of the wind overpowering that of the jet. The cocoon expands further out. Meanwhile, the jet continues to present variability, alternating phases in which it is temporarily quenched with phases in which the jet is re-energised. The latter is the consequence of fluctuating periods in which the mass accretion (and hence the powering of the jet) varies noticeably. If the power drops noticeably, so will its ram pressure, and thus, the jet may be temporarily choked. When the mass accretion increases again the jet power may, again, be enough for the jet to re-emerge. Panel c of Figure~\ref{fig1}) shows the moment at which the jet of model p0.05b has just re-surged for the third time. Its cocoon has expanded up to $y > 4 \times 10^{12}$~cm $x\sim 2\times$10$^{12}$~cm (which for this mass ratio and orbital separation is comparable to the RL of the NS).

iv) The jet variability and the quiescent/active epochs continue as the cocoon stalls. Panel d of Figure~\ref{fig1} shows how the cocoon has stalled around $\sim 2\times 10^{12}$~cm.

Jets with larger efficiency than $\eta=0.05$ will be less deflected by the bulge and the wind, moving nearly vertically across the environment with less variability and depositing less energy in the CE (although enough to remove the envelope, see \citealt{mmlcdc17}). 

The inclusion of the Coriolis and the centrifugal forces produces a shift in the alignment of the tail of the BHL bulge. In our models the bulge shifts by $\sim 4\times 10^{11}$~cm towards the centre of the stellar core of the RG (compared to when such forces are not taken into account). Since the properties of the bulge (density, pressure, bulk thickness, bulk length, and bulk height) are not affected, and $r_{\rm{in}}$ is engulfed by the BHL bulge whether such forces are accounted for or not, thus, we find that the evolution of the jet through the CE is not affected noticeably by the Coriolis nor the centrifugal force. To exclude numerical artefacts due to the domain size, we also ran an akin model of p0.05b, but in the smaller domain (model p0.05s). The evolution of the bulge, jet, and cocoon were basically the same as that in the big domain. The variable jet from model p0.05s followed the i)-iv) steps mentioned previously, with three quiescent periods, and with similar temporal mass accretion rate and specific angular momentum. 

\subsection{Mass accretion rates}
\label{sec:mdot}
In this section, we present the mass accretion rates crossing the spherical radius $r_{\rm{in}}$ ($\dot{M}_{\rm{r,in}}$) of the models where the jets are launched vertically. Before the jet is launched the quasi-steady $\dot{M}_{\rm{r,in}}$ is $\dot{M}_{\rm{r,in}}\sim$10$^{25}$~g~s$^{-1}$. If we compare it with the corresponding BHL mass accretion rate ($\dot{M}_{\rm{_{BHL}}}$), we find that the mass accretion rate is $\dot{M}_{\rm{r,in}} \sim 0.1 \dot{M}_{\rm{_{BHL}}}$ which is in accordance with previous studies \citep{beckmann18, ricker08, ricker12, macleod15a, macleod15b, macleod17a, mmlcdc17, chamandy18, gilkis19, grichener19, lcdcmm19, soker19, reichardt19a, xu19}. In our case, the diminished mass accretion rate is due to the density gradient present in the RG (see \citealt{edgar04} for further discussion).

\begin{figure}
   \includegraphics[width=0.45\textwidth]{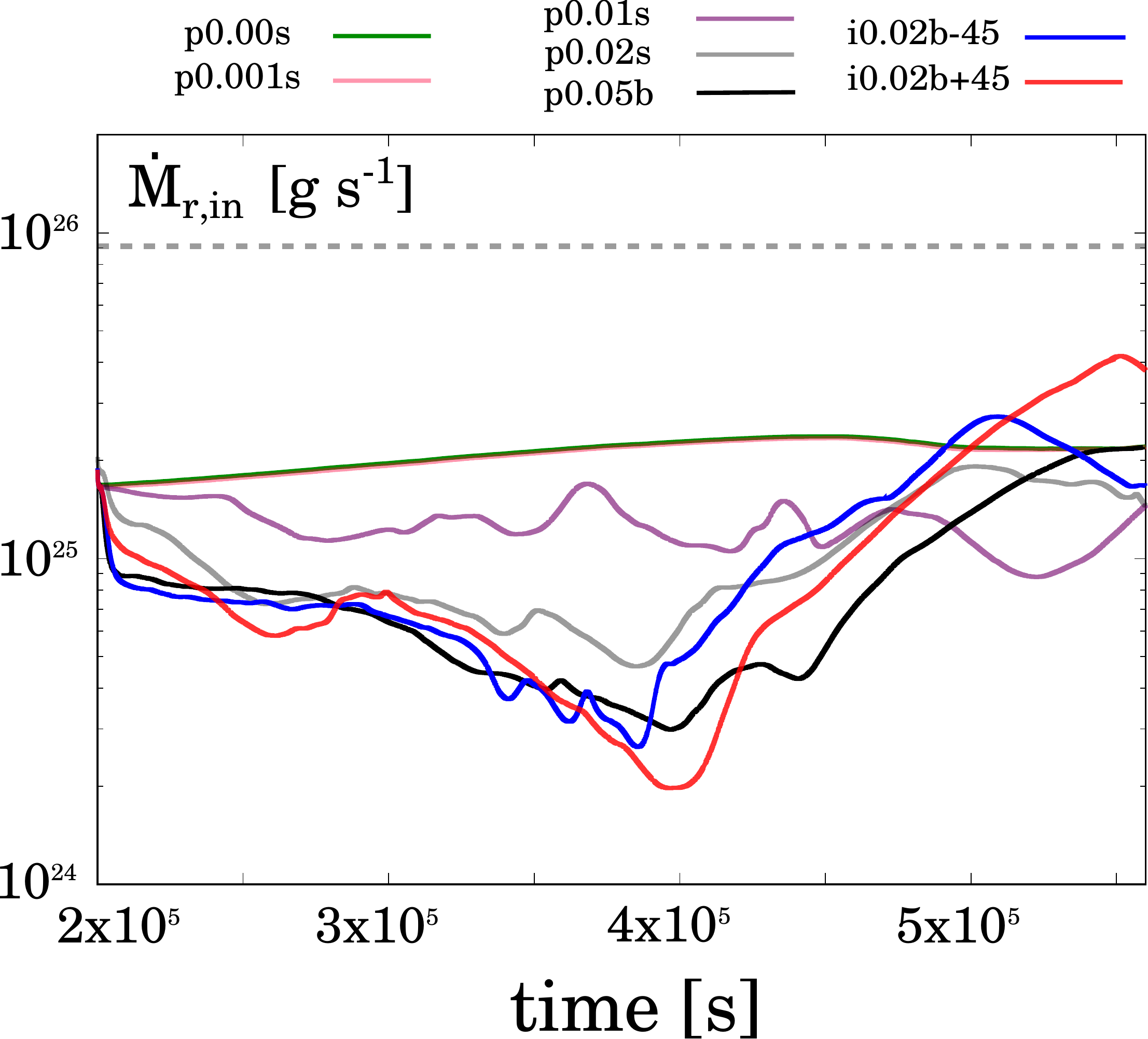}
      \caption{Mass accretion rate (in~g~s$^{-1}$) crossing the spherical inner boundary of all the models as a function of time. The grey dotted line indicates the $\dot{M}_{\rm{BHL}}$.}
   \label{fig2}      
\end{figure}

Figure~\ref{fig2} shows the $\dot{M}_{\rm{r,in}}$ of all the models as a function of time. From this figure, it is clear that if the jet is not able to drill through the bulge, the accretion rate remains basically constant ($\approx [1.7-2.1] \times 10^{25}$~g~s$^{-1}$; note that these two curves are superimposed on one another). Meanwhile, if the self-regulated jet is able to drill through the BHL bulge, the mass accretion rate will initially diminish by about a factor of two, and will then present variable behaviour. The mass accretion rate may diminish as much as an order of magnitude, but may also achieve a maximum rate $50\%$ larger than the chocked jet cases. The $\dot{M}_{\rm{r,in}}$ of the successful jet models varies between $\sim 2\times 10^{24}$~g~s$^{-1}$ and $\sim 2.1 \times 10^{25}$~g~s$^{-1}$. The obtained mass accretion rates are consistent with those from previous studies in which the r-process may ensue \citep{gilkis19, grichener19}. 

The variable behaviour of the accretion rate is due to the NJF mechanism, and is consistent with that found in \citet{lcdcmm19}. Larger $\eta$ values produce a larger NJF mechanism, and thus, larger variabilities in the $\dot{M}_{\rm{r,in}}$. The variable accretion rate behaviour of the successful jets is present throughout the entire integration time and is independent of the $\eta$ value and of the domain size (as the $\dot{M}_{\rm{r,in}}$ of models p0.05b and p0.05s are in general within 1\% of each other. 

The Eddington mass accretion rate for a NS is $\dot{M}_{\rm Edd} \sim (10^{18} {\rm{~g~s}}^{-1}) ({\rm M}/1 \msun)(0.1/\epsilon)$, i.e. $\sim 7$ orders of magnitude lower than the mass accretion rate obtained in our simulations (see Figure \ref{fig2}). The Eddington limit applies when the accretion is spherically symmetric \citep[i.e. Bondi accretion][]{bondi52}; when this is not the case, as in BHL, accretion may become hypercritical \citep{enrique08, enrique11} which is associated with jet ejection \citep[e.g.,][]{decolle12b, staff16, decolle19}.

\section{Inclination effects and disk formation}
\label{sec:incl_disk}
\subsection{Inclination effects}
\label{sec:inclination}
In this section we analyse the effects that jets launched with an inclination angle produce in the CE. As mentioned in the introduction, several processes (catastrophic mass loss events, interactions in dense cluster, Lense-Thiring and the Bardeen-Peterson effects) may interfere with the alignment of angular momentum vectors of a binary system. Thus, the spins of the stellar components and the orbital plane may be different. This will especially be likely if tidal synchronisation has not been achieved before the CE phase (as we are assuming in this study). Otherwise, the wind would likely have a velocity closer to zero (if tidal synchronisation were fully achieved) rather than (negative) Keplerian velocity, and we would need to consider Bondi accretion as opposed to BHL accretion.

In order to understand the effects that inclined jets produce in the outflow evolution, we ran simulations with the same conditions as those from the successful jet model p0.02s, with the jet launched with an inclination angle of 45$^\circ$ against or in favour of the wind (models i0.02b+45 and i0.02b-45, respectively) in a big domain. 

The global morphology of the inclined jet models and their possible disk creation is not modified by the launching angle. The representative epochs are the same as those mentioned in Section~\ref{sec:global} for the successful jets launched with no inclination angle. The main differences reside in the times at which each of the mentioned stages come into play. The jet from model i0.02b-45 (and i0.02b+45) for example, was pushed against the wind until $t\lesssim 4.0\times 10^5$~s ($t \lesssim 3.88 \times 10^5$~s) when it was realigned with the wind, and was choked in four (three) different occasions. For both inclined models, the cocoon stalled at $x\sim 2\times 10^{12}$~cm.

As for the global evolution of the system, the jet-launching angle does not significantly affect the mass accretion rate obtained at $r_{\rm{in}}$. In Figure~\ref{fig2} we see that the value of $\dot{M}_{\rm{r,in}}$ for the inclined and successful jets resembles that of the vertical successful jet models. The mass accretion rate of the inclined jets will initially diminish by about a factor of two once the jet drills through the BHL bulge and will then present a variable behaviour. Then, the inclined jet models have $\dot{M}_{\rm{r,in}} \sim 10^{25}$~g~s$^{-1}$, and compared to the vertical jet models may reach smaller or higher rates (from $\sim 2.0 \times 10^{24}$~g~s$^{-1}$ up to $4.5\times 10^{25}$~g~s$^{-1}$). The mass accretion rate variable behaviour of the inclined jets is present throughout the entire integration time.

\subsection{Disk formation}
\label{sec:disk}
In this section we discuss whether the accreted material may form an accretion disk inside $r_{\rm{in}}$. We note that the presence of an accretion disk is a key ingredient in order to launch a jet as the accretion may be accompanied by polar outflows. The introduction of a density gradient in the CE profile (as well as Coriolis and centrifugal forces) allows for angular momentum to be introduced to the flow, which in turn opens the possibility for the formation of a disk around the embedded companion \citep{murguia17}. We must note that the most well supported energy extraction mechanisms in order to launch jets involve the presence of an accretion disk around the compact object. It can either be due to energy extraction from a rotating accretion disk with a frozen magnetic field \citep[][]{bz77,bp82}{\footnote{We include \citet[][]{bz77} as, even though it is designed for a Kerr BH, one could still extract energy from a rapidly rotating NS.}}, or the extraction from neutrino annihilation produced in an accretion disk \citep[][]{pwf99}. Since the triggering of the jet is far from the scope of this study, we limit our analysis to the possible formation of an accretion disk once it has crossed the inner boundary.

\begin{figure}
   \includegraphics[width=0.43\textwidth]{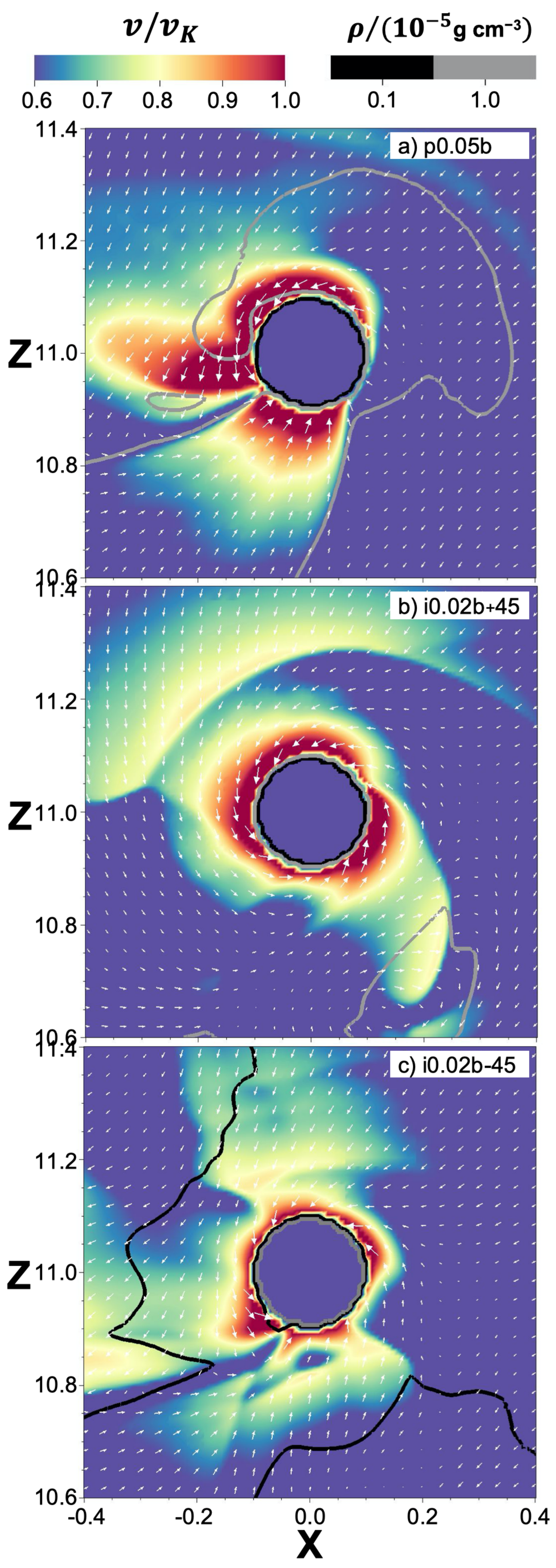}
   \caption{Equatorial velocity magnitude map (normalised to v$_{\rm{k}}$=4.3$\times 10^{7}$~cm~s$^{-1}$) at 5.51$ \times 10^{5}$~s for models p0.05b (top panel), i0.02b+45 (middle), and i0.02b-45 (bottom). The velocity vectors range from 0.53 to 1 v$_{\rm{K}}$. The black, grey, and white isocontour lines correspond to density values equal to 0.6, 0.8, and 1 (in units of $10^{-5}$g~cm$^{-3}$). The axis are in units of $10^{12}$~cm.}   
   \label{fig3}
\end{figure}

In order to analyse the possible formation of a disk we study whether the material around $r_{\rm{in}}$ has tangential velocities which are comparable to $v_{\rm{k}}$. The Keplerian velocity at $r_{\rm{in}}$ is $v_{\rm{k}}=4.3\times 10^{7}$~cm~s$^{-1}$. In Figure~\ref{fig3} we present the equatorial velocity magnitude map (normalised  with respect to $v_{\rm{k}}$) for models p0.05b, i0.02b+45, and i0.02b-45 at 5.51$\times 10^{5}$~s (the velocity field, and density isocontours are also indicated). For the three models, a large fraction of the material that surrounds $r_{\rm{in}}$ have both a tangential component and $v\simeq v_{\rm{k}}$ around $r_{\rm{in}}$, suggesting that independently of the launching angle of the jet, a disk may form around $r_{\rm{in}}$ (or inside of it).

\begin{figure}
   \includegraphics[width=0.45\textwidth]{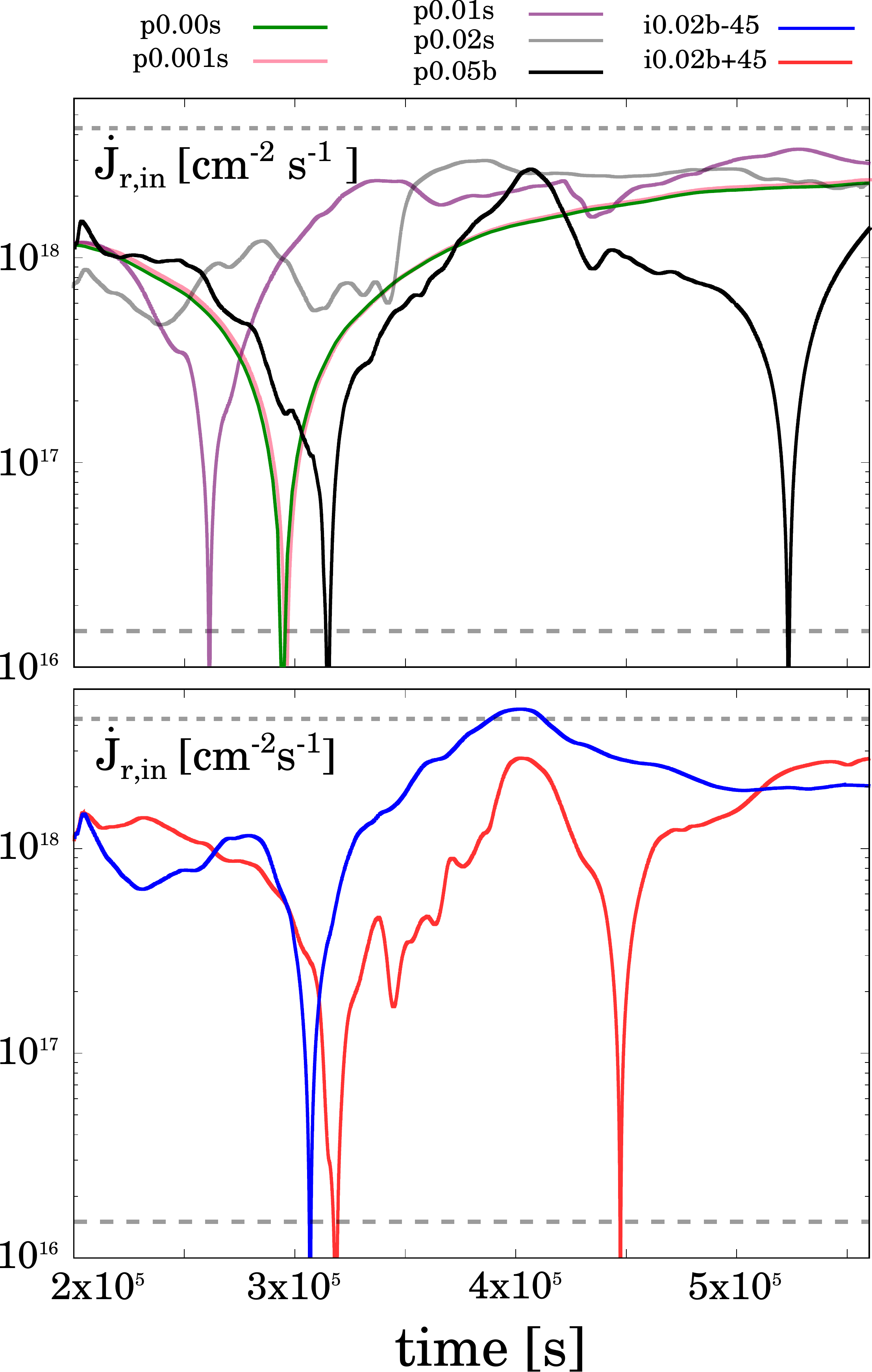}
      \caption{Absolute specific angular momentum (in cm$^{-2}$~s$^{-1}$) crossing the spherical inner boundary for the vertical and inclined-jet models (top and bottom panels, respectively) as a function of time. The upper grey dotted line and lower grey dashed line correspond to the $\dot{J}_{\rm{crit, in}}$ and $\dot{J}_{\rm{crit, NS}}$, respectively.}
   \label{fig4}
\end{figure}

To form an accretion disk, the specific angular momentum must be equal or larger than the critical value $\dot{J}_{\rm{crit}} = \sqrt{\rm{G M R}}$ (being $M$ the mass of the NS, and $R$ the radius of interest). Thus, the specific angular momentum to form a disk at $r_{\rm{in}}$ is $\dot{J}_{\rm{crit, in}}=4.30\times 10^{18}$~cm$^{-2}$~s$^{-1}$, and at the surface of the NS is $\dot{J}_{\rm{crit, NS}}=1.51\times 10^{16}$~cm$^{-2}$~s$^{-1}$. Figure~\ref{fig4} shows the temporal evolution of the absolute value of the specific angular momentum that crosses the inner boundary{\footnote{We must note that, as for the mass accretion rate, for the specific angular momentum calculation we take into account all of the material that crosses the spherical inner boundary and not only that at the equatorial plane shown in Figure~\ref{fig3}.}} ($\left|{\dot{J}_{\rm{r,in}}}\right|$) for the vertical and inclined-jet models (upper and lower panels, respectively; the values of $\dot{J}_{\rm{crit, in}}$ and $\dot{J}_{\rm{crit, NS}}$ are also indicated).

The specific angular momentum of the unsuccessful (or quenched) jet models is $\left|{\dot{J}_{\rm{r,in}}}\right| \sim 1$-$2\times 10^{18}$~cm$^{-2}$~s$^{-1}$ ($\sim 2$-$3\times 10^{18}$~cm$^{-2}$~s$^{-1}$) and has one rapid dip-rise at $\sim 3.0 \times 10^{5}$~s ($\sim 2.6 \times 10^{5}$~s). The specific angular momentum of the successful jets vertically launched ($\left|{\dot{J}_{\rm{r,in}}}\right| \sim$2-4$\times 10^{18}$~cm$^{-2}$~s$^{-1}$) reaches larger values that that of the non-successful jets. The inclined jet models values are similar to the vertical jet models but grasps even larger values (model i0.02b-45, for example, reaches $\sim 4\times 10^{18}$~cm$^{-2}$~s$^{-1}$ at $4 \times 10^{5}$~s). The sudden dip-rise corresponds to the moment in which the overall direction of the material that crosses $r_{\rm{in}}$ changes. Such spin changes are produced when the self-regulated jet are noticeably reduced (or choked) and re-surge. In some models the material surrounding $r_{\rm{in}}$ may present none or more than one spin change (see for example models p0.02s; and p0.05b, i0.02b+45, respectively).  

Unlike previous studies which state that the formation of an accretion disk in the CE is rare or transitory \citep{murguia17}, our results suggest that accretion disk will form independently of the ram pressure, or the inclination angle of the jet. Regardless of the parameters used in our study, and the domain size{\footnote{The $\left|{\dot{J}_{\rm{r,in}}}\right|$ of models p0.05s and p0.05b are in general within $\pm$50\% of each other.}}, the $\left|{\dot{J}_{\rm{r,in}}}\right|$ is within a fraction of $\dot{J}_{\rm{crit, in}}$ (or two orders of magnitude above $\dot{J}_{\rm{crit, NS}}$). The specific angular momentum obtained with our simulations is consistent with the one obtained by \citet{macleod17a}. Hence, we infer that the material that crosses $r_{\rm{in}}$ will form an accretion disk around the NS (since $\left|{\dot{J}_{\rm{r,in}}}\right| \geq \dot{J}_{\rm{crit, NS}}$). Assuming angular momentum conservation, from the values we obtain we find that a disk will be formed at a radius of about $\gtrsim 10^{10}$ cm. In some cases, as that for model i0.02b-45, a transitory disk may form around $r_{\rm{in}}$. Also, we find that there is a mild anti-correlation between the mass accretion rate and the specific angular momentum. This is due to the fact that when high angular momentum is present the material will take longer to accrete.

Since the binding energy of the envelope is $\sim 10^{49}$erg~s$^{-1}$ \citep{lcdcmm19} and the jet luminosity in our study is of order $\sim 10^{43}$erg~s$^{-1}$, the jet would require approximately $10^{6}$~s to stop the CE phase. Thus, we estimate that the mass of the disk would be approximately 10$^{-2}$~M$_\odot$ which is in accordance with previous studies \citep[see, e.g.,][for the case without a jet]{staff16}.

Finally, we notice that the equations of hydrodynamics can be re-scaled by considering proper relations between the physical quantities (the domain size, the velocity, and the integration time). In particular, if we keep the mass of the compact object fixed, the results of the simulations will be valid if we consider the local Keplerian velocity ($v_{\rm{k,l}}$) for the velocity of the wind, and if the domain is rescaled by a factor $\propto 1/v_{\rm{k,l}}^2$. Thus, the results obtained can be generalised to get a disk radius r$_{\rm disk} \sim (10^{10}~{\rm cm}) (a/10^{13}{\rm cm})^{1/2}$ (with $a$ the orbital separation).

\section{Conclusions}
\label{sec:con}
Jets launched from a CO in a CE may play a key role in the evolution of the system, and may be an efficient removal channel for the material of the CE. Thus, in this paper, we follow the interaction between the material that is accreted by NS, the incoming wind material (due to the orbital motion of the NS within the CE and that we are in the comoving frame), and the jet which is launched from the NS (including the formed cocoon). Apart from the global dynamical evolution of a jet (and its correspondent cocoon) through the CE, we examine the effects produced by inclined jets, as well as the possible formation of a disk. 

We follow a set of 3D-HD simulations in which the jet may be launched with different luminosities and launching angles. The luminosity of the jet is self-regulated by the mass accretion rate and an efficiency $\eta$. The Coriolis and centrifugal forces are included, and we follow the models for up to approximately a week. We focus on the the mass- and specific-angular-momentum accretion rates in order to analyse the possible formation of an accretion disk. The simulation domain covers approximately $\sim$10\% of the CE of a 20 M$_\odot$ RG, and we model only the top half of the system. We assume that the jet is launched from a $1.4$ M$_\odot$ NS orbiting around the centre of mass of the RG with Keplerian velocity. The simulations are done in the reference frame co-rotating with the NS.

Independently of the luminosity or the inclination angle of the jet, the specific angular momentum is above the critical value in order to form an accretion disk at a radius of about $\gtrsim 10^{10}$ cm around the NS. In some cases, we find that a transitory disk may form around $\gtrsim 10^{11}$ cm. The final mass of the disk will be of order 0.04-0.06 M$_\odot$. 

The general evolution and morphology of all the jets (inclined or not) which are able to break out of the BHL bulge mainly follows four stages. These are: i) Depending on the $\eta$ efficiency value the jet may be able to drill through a previously formed BHL bulge. Due to the NJF mechanism the jet will present variable behaviour (in power, size, and direction) throughout this and the next stages. The environment and the jet material are shocked by the forward and reverse jet shock respectively and form a cocoon which expands smoothly over the CE. ii) The successful jet is pushed against the wind due to the pressure of the bulge. iii) The jets are realigned with the direction of the incoming wind (due to the ram pressure of the wind overpowering that of the jet). The jets may present quiescent and active epochs. This is due to the alternation of periods of mass accretion in which the ram pressure of the jet may be quenched and may later re-surge. iv) The jet quiescent and active epochs continue as the cocoon stalls. The $\eta$ and inclination angle will affect the moment in which they come into play, as well as the amount of jet choking and re-surging epochs. The latter depend on the efficiency of the accretion to ejection process (i.e. $\eta$) and on the inclination angle of the jet. We must note that the inclusion of the Coriolis and centrifugal forces does not bring new important effects as the overall results are consistent with prior studies of jets launched from NS within the CE but where the extra forces were not considered \citep{mmlcdc17, lcdcmm19}. 
 
For the choked jets the mass accretion rate is $\sim 20\%$ of the BHL accretion rate. The successful jets have mass accretion rates which oscillate close to $\sim 2$-$40$\% of the BHL rate. In our case the low mass accretion rates obtained are due the presence of density gradient in the CE profile, and are consistent with previous BHL accretion numerical studies. 

The inclination of the jet modifies the time at which the previously mentioned stages of the global evolution (i-iv) will come into play. Also it may produce a larger range of mass accretion rate. Nevertheless, we find that inclined jets do not substantially affect the global evolution nor the mass-accretion rate when compared to vertically-launched jets. 


There is a mild anti-correlation between the mass accretion rate and the specific angular momentum. Also, the angular momentum that crosses the inner boundary may change its overall sign randomly with time, we expect to get disks formed by rings with alternate rotation directions. This will lead to shearing between different layers of the disk, which in turn, will produce heating and loss of angular momentum in these regions. As the shearing regions approach the NS, the jet will vary drastically its power. 

The variable behaviour of the jets and the active/dormant epochs may produce vortices in the CE. The latter effect, along with the recombination energy \citep{nandez16, ivanova13}, may play a role in terminating the CE phase \citep{schreier19}. 

Our simulations assume that the companion is a NS (engulfed in a massive RG). Nevertheless, our results can be generalised to WDs or main sequence stars as the jet dynamics is mainly governed by its ram pressure ($P_{\rm ram}=\rho_j v_j^2 \propto \dot{M_j} v_j \propto \eta \dot{M_a} v_j)$. Assuming that the velocity of the jet is the escape velocity from the companion ($v_{\rm esc}\propto \sqrt{M/r}$), and considering the minimum efficiency to get a successful jet (in our case $\eta =\eta_{\rm NS} \gtrsim 0.02$), we obtain that for a WD engulfed in the CE, the mass-accretion-to-jet-ejection process needs to be very efficient ($\gtrsim$ 25 times more than for the NS case, i.e. $\eta\gtrsim \eta_{\rm NS} \sqrt{R_{\rm NS}/R_{\rm WD}}\gtrsim 0.5$) in order to get a successful jet in the outer layers of the CE. For main sequence stars, there is not enough energy reservoir to launch a successful jet. This is a consequence of the lower binding energy ($\sim 10^5$ times less) available in less compact stars.

We note that we do not follow the evolution of the full system (i.e. the jets in both hemispheres) and that we are neglecting the effects due to the opposite jet and cocoon which will not be symmetric for the inclined jet models. Future simulations and studies are necessary in order to fully understand the latter effects, as well as the the disk formation, stability, and shearing effects; the accretion/ejection process; and the possible termination of the CE phase via jets within CEs.

\section*{Acknowledgements}
D.L.C. is supported by C\'atedras CONACyT at the Instituto de Astronom\'ia (UNAM). We acknowledge the support from the Miztli-UNAM supercomputer (projects LANCAD-UNAM-DGTIC-321 and LANCAD-UNAM-DGTIC-281 for the assigned computational time in which the simulations were performed. F.D.C. thanks the UNAM-PAPIIT grant AG100820.


\bsp	
\label{lastpage}

\begin{thebibliography}{99}
\bibitem[\protect\citeauthoryear{Abbott et al.}{2016}]{abbott16} Abbott B.~P., et al., 2016, PhRvL, 116, 241103 
\bibitem[\protect\citeauthoryear{Armitage \& Livio}{2000}]{armitage00} Armitage P.~J., Livio M., 2000, \apj, 532, 540
\bibitem[\protect\citeauthoryear{Banerjee}{2018}]{banerjee18} Banerjee S., 2018, MNRAS, 473, 909
\bibitem[\protect\citeauthoryear{Bardeen \& Petterson}{1975}]{bp75} Bardeen J.~M., Petterson J.~A., 1975, ApJL, 195, L65
\bibitem[Beckmann et al.(2018)]{beckmann18} Beckmann R.~S., Slyz A., Devriendt J., 2018, \mnras, 478, 995
\bibitem[Blandford \& Payne(1982)]{bp82} Blandford, R.~D., \& Payne, D.~G.\ 1982, \mnras, 199, 883  
\bibitem[Blandford \& Znajek(1977)]{bz77} Blandford, R.~D., \& Znajek, R.~L.\ 1977, \mnras, 179, 433
\bibitem[\protect\citeauthoryear{Bodenheimer \& Taam}{1984}]{bodenheimer84} Bodenheimer P., Taam R.~E., 1984, ApJ, 280, 771
\bibitem[Bondi(1952)]{bondi52} Bondi, H.\ 1952, \mnras, 112, 195 
\bibitem[\protect\citeauthoryear{Bondi \& Hoyle}{1944}]{bh44} Bondi H., Hoyle F., 1944, \mnras, 104, 273 
\bibitem[\protect\citeauthoryear{Brown et al.}{2007}]{Brown2007} Brown G.~E., Lee C.-H., Moreno M{\'e}ndez E., 2007, ApJL, 671, L41
\bibitem[\protect\citeauthoryear{Chamandy, et al.}{2019a}]{chamandy19a} Chamandy L., Blackman E.~G., Frank A., Carroll-Nellenback J., Zou Y., Tu Y., 2019, MNRAS, 490, 3727
\bibitem[\protect\citeauthoryear{Chamandy, et al.}{2018}]{chamandy18} Chamandy L., et al., 2018, MNRAS, 480, 1898
\bibitem[\protect\citeauthoryear{Chamandy, et al.}{2019b}]{chamandy19b} Chamandy L., Tu Y., Blackman E.~G., Carroll-Nellenback J., Frank A., Liu B., Nordhaus J., 2019, MNRAS, 486, 1070
\bibitem[\protect\citeauthoryear{Chevalier}{2012}]{chevalier12} Chevalier R.~A., 2012, ApJL, 752, L2
\bibitem[\protect\citeauthoryear{De Colle et al.}{2012a}]{decolle12a} De Colle F., Granot J., L{\'o}pez-C{\'a}mara D., Ramirez-Ruiz E., 2012a, ApJ, 746, 122 
\bibitem[De Colle et al.(2012)]{decolle12b} De Colle, F., Guillochon, J., Naiman, J., et al.\ 2012, \apj, 760, 103
\bibitem[De Colle \& Lu(2019)]{decolle19} De Colle, F., \& Lu, W.\ 2019, arXiv e-prints, arXiv:1911.01442
\bibitem[\protect\citeauthoryear{De Marco, et al.}{2011}]{demarco11} De Marco O., Passy J.-C., Moe M., Herwig F., Mac Low M.-M., Paxton B., 2011, MNRAS, 411, 2277
\bibitem[\protect\citeauthoryear{De Marco et al.}{2003}]{demarco03} De Marco O., Sandquist E.~L., Mac Low M.-M., Herwig F., Taam R.~E., 2003, RMxAC, 15, 34
\bibitem[\protect\citeauthoryear{Edgar}{2004}]{edgar04} Edgar R., 2004, NewAR, 48, 843
\bibitem[\protect\citeauthoryear{El Mellah, Sundqvist \& Keppens}{2019}]{elmellah19} El Mellah I., Sundqvist J.~O., Keppens R., 2019, A\&A, 622, L3
\bibitem[\protect\citeauthoryear{Fragos, et al.}{2019}]{fragos19} Fragos T., Andrews J.~J., Ramirez-Ruiz E., Meynet G., Kalogera V., Taam R.~E., Zezas A., 2019, ApJL, 883, L45
\bibitem[\protect\citeauthoryear{Gilkis, Soker \& Kashi}{2019}]{gilkis19} Gilkis A., Soker N., Kashi A., 2019, \mnras, 482, 4233
\bibitem[\protect\citeauthoryear{Glanz \& Perets}{2018}]{glanz18} Glanz H., Perets H.~B., 2018, MNRAS, 478, L12
\bibitem[\protect\citeauthoryear{Glanz \& Perets}{2020}]{glanz20} Glanz H., Perets H.~B., 2020, arXiv, arXiv:2004.00020
\bibitem[Grichener et al.(2018)]{grichener18} Grichener A., Sabach E., Soker N., 2018, MNRAS, 478, 1818
\bibitem[\protect\citeauthoryear{Grichener \& Soker}{2019}]{grichener19} Grichener A., Soker N., 2019, ApJ, 878, 24
\bibitem[Han et al.(1994)]{han94} Han, Z., Podsiadlowski, P., \& Eggleton, P.~P.\ 1994, \mnras, 270, 121
\bibitem[\protect\citeauthoryear{Hillel, Schreier \& Soker}{2017}]{hillel17} Hillel S., Schreier R., Soker N., 2017, MNRAS, 471, 3456
\bibitem[\protect\citeauthoryear{Iaconi \& De Marco}{2019}]{iaconi19a} Iaconi R., De Marco O., 2019, MNRAS, 490, 2550
\bibitem[Iaconi et al.(2018)]{iaconi18} Iaconi, R., De Marco O., Passy J.-C., Staff J., 2018, \mnras, 477, 2349
\bibitem[\protect\citeauthoryear{Iaconi, et al.}{2019}]{iaconi19b} Iaconi R., Maeda K., De Marco O., Nozawa T., Reichardt T., 2019, MNRAS, 489, 3334
\bibitem[Iaconi et al.(2017)]{iaconi17} Iaconi, R., Reichardt, T., Staff, J., et al.\ 2017, \mnras, 464, 4028
\bibitem[Iben \& Livio(1993)]{iben93} Iben, I., Jr., \& Livio, M.\ 1993, \pasp, 105, 1373 
\bibitem[\protect\citeauthoryear{Igoshev, Perets \& Michaely}{2019}]{igoshev19} Igoshev A.~P., Perets H.~B., Michaely E., 2019, arXiv, arXiv:1907.10068 
\bibitem[Ivanova \& Chaichenets(2011)]{ivanova11} Ivanova, N., \& Chaichenets, S.\ 2011, \apjl, 731, L36 
\bibitem[\protect\citeauthoryear{Ivanova \& Nandez}{2016}]{ivanova16} Ivanova N., Nandez J.~L.~A., 2016, MNRAS, 462, 362
\bibitem[Ivanova \& Podsiadlowski(2003)]{ivanova03} Ivanova, N., \& Podsiadlowski, P.\ 2003, From Twilight to Highlight: The Physics of Supernovae, 19
\bibitem[\protect\citeauthoryear{Ivanova et al.}{2013}]{ivanova13} Ivanova N., et al., 2013, A\&ARv, 21, 59 
\bibitem[\protect\citeauthoryear{Jones}{2020}]{jones20} Jones D., 2020, arXiv, arXiv:2001.03337 
\bibitem[Kuruwita et al.(2016)]{kuruwita16} Kuruwita, R.~L., Staff, J., \& De Marco, O.\ 2016, \mnras, 461, 486
\bibitem[Lense \& Thirring (1918) ]{lt18} Lense,~J., \& Thirring,~H., 1918, Physikalische Zeitschrift, 19, 156 
\bibitem[Livio \& Soker(1988)]{livio88} Livio, M., \& Soker, N.\ 1988, \apj, 329, 764
\bibitem[\protect\citeauthoryear{L{\'o}pez-C{\'a}mara, De Colle \& Moreno M{\'e}ndez}{2019}]{lcdcmm19} L{\'o}pez-C{\'a}mara D., De Colle F., Moreno M{\'e}ndez E., 2019, MNRAS, 482, 3646
\bibitem[\protect\citeauthoryear{Lombardi, et al.}{2006}]{lombardi06} Lombardi J.~C., Proulx Z.~F., Dooley K.~L., Theriault E.~M., Ivanova N., Rasio F.~A., 2006, ApJ, 640, 441
\bibitem[\protect\citeauthoryear{MacLeod \& Ramirez-Ruiz}{2015a}]{macleod15a} MacLeod M., Ramirez-Ruiz E., 2015, ApJ, 803, 41 
\bibitem[\protect\citeauthoryear{MacLeod \& Ramirez-Ruiz}{2015b}]{macleod15b} MacLeod M., Ramirez-Ruiz E., 2015, ApJL, 798, L19
\bibitem[\protect\citeauthoryear{MacLeod, et al.}{2017a}]{macleod17a} MacLeod M., Antoni A., Murguia-Berthier A., Macias P., Ramirez-Ruiz E., 2017, ApJ, 838, 56
\bibitem[\protect\citeauthoryear{MacLeod, et al.}{2017b}]{macleod17b} MacLeod M., Macias P., Ramirez-Ruiz E., Grindlay J., Batta A., Montes G., 2017, ApJ, 835, 282
\bibitem[\protect\citeauthoryear{Meyer \& Meyer-Hofmeister}{1979}]{meyer79} Meyer F., Meyer-Hofmeister E., 1979, A\&A, 78, 167
\bibitem[\protect\citeauthoryear{Mohamed \& Podsiadlowski}{2007}]{mp07} Mohamed S., Podsiadlowski P., 2007, ASPC, 372, 397, ASPC..372
\bibitem[Moreno M{\'e}ndez et al.(2017)]{mmlcdc17} Moreno M{\'e}ndez, E., L{\'o}pez-C{\'a}mara, D., \& De Colle, F.\ 2017, MNRAS, 470, 2929 
\bibitem[Moreno M{\'e}ndez et al.(2008)]{enrique08} Moreno M{\'e}ndez, E., Brown, G.~E., Lee, C.-H., \& Park, I.~H.\ 2008, \apjl, 689, L9 
\bibitem[Moreno M{\'e}ndez(2011)]{enrique11} Moreno M{\'e}ndez, E.\ 2011, \mnras, 413, 183 
\bibitem[\protect\citeauthoryear{Moreno M{\'e}ndez \& Cantiello}{2016}]{enrique16} Moreno M{\'e}ndez E., Cantiello M., 2016, NewA, 44, 58
\bibitem[\protect\citeauthoryear{Murguia-Berthier, et al.}{2017}]{murguia17} Murguia-Berthier A., MacLeod M., Ramirez-Ruiz E., Antoni A., Macias P., 2017, ApJ, 845, 173
\bibitem[Nandez et al.(2014)]{nandez14} Nandez, J.~L.~A., Ivanova, N., \& Lombardi, J.~C., Jr.\ 2014, \apj, 786, 39
\bibitem[\protect\citeauthoryear{Nandez, Ivanova \& Lombardi}{2015}]{nandez15} Nandez J.~L.~A., Ivanova N., Lombardi J.~C.~J., 2015, MNRAS, 450, L39
\bibitem[\protect\citeauthoryear{Nandez \& Ivanova}{2016}]{nandez16} Nandez J.~L.~A., Ivanova N., 2016, MNRAS, 460, 3992
\bibitem[Nelemans et al.(2000)]{nelemans00} Nelemans, G., Verbunt, F., Yungelson, L.~R., \& Portegies Zwart, S.~F.\ 2000, \aap, 360, 1011 
\bibitem[Ohlmann et al.(2016a)]{ohlmann16a} Ohlmann, S.~T., R{\"o}pke, F.~K., Pakmor, R., \& Springel, V.\ 2016, \apjl, 816, L9 
\bibitem[Ohlmann et al.(2016b)]{ohlmann16b} Ohlmann, S.~T., R{\"o}pke, F.~K., Pakmor, R., Springel, V., \& M{\"u}ller, E.\ 2016, \mnras, 462, L121 
\bibitem[\protect\citeauthoryear{Ohlmann, et al.}{2017}]{ohlmann17} Ohlmann S.~T., R{\"o}pke F.~K., Pakmor R., Springel V., 2017, A\&A, 599, A5
\bibitem[\protect\citeauthoryear{Papish, Soker, \& Bukay}{2015}]{papish15} Papish O., Soker N., Bukay I., 2015, MNRAS, 449, 288 
\bibitem[\protect\citeauthoryear{Passy, et al.}{2011}]{passy11} Passy J.-C., et al., 2011, ASPC, 447, 107, ASPC..447
\bibitem[\protect\citeauthoryear{Passy et al.}{2012}]{passy12} Passy J.-C., et al., 2012, ApJ, 744, 52 
\bibitem[Paczynski(1976)]{paczynski76} Paczynski, B.\ 1976, Structure and Evolution of Close Binary Systems, 73, 75 
\bibitem[\protect\citeauthoryear{Pejcha, Metzger \& Tomida}{2016}]{pejcha16} Pejcha O., Metzger B.~D., Tomida K., 2016, MNRAS, 455, 4351
\bibitem[Popham et al.(1999)]{pwf99} Popham, R., Woosley, S.~E., \& Fryer, C.\ 1999, \apj, 518, 356 
\bibitem[\protect\citeauthoryear{Postnov \& Yungelson}{2014}]{py14} Postnov K.~A., Yungelson L.~R., 2014, LRR, 17, 3
\bibitem[\protect\citeauthoryear{Prust \& Chang}{2019}]{prust19} Prust L.~J., Chang P., 2019, MNRAS, 486, 5809
\bibitem[\protect\citeauthoryear{Prust}{2020}]{prust20} Prust L.~J., 2020, arXiv, arXiv:2002.04287 
\bibitem[\protect\citeauthoryear{Ramirez-Ruiz \& Lee}{2009}]{rr09} Ramirez-Ruiz E., Lee W., 2009, Nature, 460, 1091
\bibitem[\protect\citeauthoryear{Rasio \& Shapiro}{1991}]{rasio91} Rasio F.~A., Shapiro S.~L., 1991, ApJ, 377, 559
\bibitem[\protect\citeauthoryear{Rasio \& Livio}{1996}]{rasio96} Rasio F.~A., Livio M., 1996, ApJ, 471, 366 
\bibitem[\protect\citeauthoryear{Reichardt, et al.}{2019a}]{reichardt19a} Reichardt T.~A., De Marco O., Iaconi R., Tout C.~A., Price D.~J., 2019, MNRAS, 484, 631
\bibitem[\protect\citeauthoryear{Reichardt, et al.}{2019b}]{reichardt19b} Reichardt T., De Marco O., Iaconi R., Price D., 2019, arXiv, arXiv:1911.02759 
\bibitem[Reg{\H o}s \& Tout(1995)]{regos95} Reg{\H o}s, E., \& Tout, C.~A.\ 1995, \mnras, 273, 146 
\bibitem[\protect\citeauthoryear{Ricker \& Taam}{2008}]{ricker08} Ricker P.~M., Taam R.~E., 2008, ApJ, 672, L41 
\bibitem[Ricker \& Taam(2012)]{ricker12} Ricker, P.~M., \& Taam, R.~E.\ 2012, \apj, 746, 74 
\bibitem[\protect\citeauthoryear{Sabach, et al.}{2017}]{sabach17} Sabach E., Hillel S., Schreier R., Soker N., 2017, MNRAS, 472, 4361
\bibitem[\protect\citeauthoryear{Sandquist, Taam, \& Burkert}{2000}]{sandquist00} Sandquist E.~L., Taam R.~E., Burkert A., 2000, ApJ, 533, 984 
\bibitem[\protect\citeauthoryear{Sandquist et al.}{1998}]{sandquist98} Sandquist E.~L., Taam R.~E., Chen X., Bodenheimer P., Burkert A., 1998, ApJ, 500, 909 
\bibitem[\protect\citeauthoryear{Schreier, Hillel \& Soker}{2019}]{schreier19} Schreier R., Hillel S., Soker N., 2019, MNRAS, 490, 4748
\bibitem[\protect\citeauthoryear{Shiber, Schreier \& Soker}{2016}]{shiber16} Shiber S., Schreier R., Soker N., 2016, RAA, 16, 117
\bibitem[Shiber et al.(2017)]{shiber17} Shiber, S., Kashi, A., \& Soker, N.\ 2017, \mnras, 465, L54
\bibitem[Shiber \& Soker(2018)]{shiber18} Shiber, S., \& Soker, N.\ 2018, \mnras, 477, 2584
\bibitem[Shiber et al.(2019)]{shiber19} Shiber S., Iaconi R., De Marco O., Soker N., 2019, \mnras, 488, 5615
\bibitem[Soker(2004)]{soker04} Soker, N.\ 2004, \na, 9, 399
\bibitem[\protect\citeauthoryear{Soker}{2013a}]{soker13a} Soker N., 2013, NewA, 18, 18
\bibitem[\protect\citeauthoryear{Soker, et al.}{2013b}]{soker13b} Soker N., Akashi M., Gilkis A., Hillel S., Papish O., Refaelovich M., Tsebrenko D., 2013, AN, 334, 402
\bibitem[\protect\citeauthoryear{Soker}{2014}]{soker14} Soker N., 2014, arXiv, arXiv:1404.5234 
\bibitem[Soker(2016)]{soker16} Soker, N.\ 2016, \nar, 75, 1
\bibitem[\protect\citeauthoryear{Soker \& Gilkis}{2018}]{soker18} Soker N., Gilkis A., 2018, MNRAS, 475, 1198
\bibitem[\protect\citeauthoryear{Soker, Grichener \& Gilkis}{2019}]{soker19} Soker N., Grichener A., Gilkis A., 2019, \mnras, 484, 4972
\bibitem[\protect\citeauthoryear{Spruit \& Phinney}{1998}]{spruit98} Spruit H., Phinney E.~S., 1998, Natur, 393, 139
\bibitem[\protect\citeauthoryear{Staff, et al.}{2016}]{staff16} Staff J.~E., De Marco O., Macdonald D., Galaviz P., Passy J.-C., Iaconi R., Low M.-M.~M., 2016, MNRAS, 455, 3511
\bibitem[\protect\citeauthoryear{Taam, Bodenheimer \& Ostriker}{1978}]{taam78} Taam R.~E., Bodenheimer P., Ostriker J.~P., 1978, ApJ, 222, 269
\bibitem[\protect\citeauthoryear{Taam \& Ricker}{2010}]{taam10} Taam R.~E., Ricker P.~M., 2010, NewAR, 54, 65
\bibitem[\protect\citeauthoryear{Terman, Taam, \& Hernquist}{1994}]{terman94} Terman J.~L., Taam R.~E., Hernquist L., 1994, ApJ, 422, 729 
\bibitem[\protect\citeauthoryear{van den Heuvel}{1976}]{heuvel76} van den Heuvel E.~P.~J., 1976, IAUS, 73, 35
\bibitem[\protect\citeauthoryear{Vigna-G{\'o}mez, et al.}{2020}]{vg20} Vigna-G{\'o}mez A., et al., 2020, arXiv, arXiv:2001.09829 
\bibitem[Voss \& Tauris(2003)]{voss03} Voss, R., \& Tauris, T.~M.\ 2003, \mnras, 342, 1169 
\bibitem[\protect\citeauthoryear{Wongwathanarat, Janka \& M{\"u}ller}{2013}]{wongwa13} Wongwathanarat A., Janka H.-T., M{\"u}ller E., 2013, A\&A, 552, A126
\bibitem[\protect\citeauthoryear{Xu \& Stone}{2019}]{xu19} Xu W., Stone J.~M., 2019, MNRAS, 488, 5162
\end{thebibliography}
\end{document}